\documentclass[conference]{IEEEtran}

\usepackage{booktabs} 
\usepackage{xargs} 
\usepackage{xcolor}  
\usepackage[colorinlistoftodos,prependcaption,textsize=normalsize]{todonotes}

\usepackage{cite}

\usepackage{graphicx}

\usepackage{amsmath}
\usepackage{algorithmic}

\usepackage{array}

\usepackage{url}

\newcommand{\code}{\texttt}

\begin{document}
\title{Shrink or Substitute: Handling Process Failures in HPC Systems using In-situ Recovery}

\IEEEoverridecommandlockouts

\author
{
\IEEEauthorblockN{Rizwan A. Ashraf, Saurabh Hukerikar and Christian Engelmann}
\IEEEauthorblockA{Computer Science and Mathematics Division,\\
Oak Ridge National Laboratory,\\
Oak Ridge, TN 37831, USA.\\
Email: \{ashrafra, hukerikarsr, engelmannc\}@ornl.gov}
\thanks{This manuscript has been authored by UT-Battelle, LLC under Contract No. DE-AC05-00OR22725 with the U.S. Department of Energy. The United States Government retains and the publisher, by accepting the article for publication, acknowledges that the United States Government retains a non-exclusive, paid-up, irrevocable, worldwide license to publish or reproduce the published form of this manuscript, or allow others to do so, for United States Government purposes. The Department of Energy will provide public access to these results of federally sponsored research in accordance with the DOE Public Access Plan (http://energy.gov/downloads/doe-public-access-plan).}}

\maketitle

\begin{abstract}
Efficient utilization of today's high-performance computing (HPC) systems with complex hardware and software components requires that the HPC applications are designed to tolerate process failures at runtime.
With low mean-time-to-failure (MTTF) of current and future HPC systems, long running simulations on these systems require capabilities for gracefully handling process failures by the applications themselves.
In this paper, we explore the use of fault tolerance extensions to Message Passing Interface (MPI) called user-level failure mitigation (ULFM) for handling process failures without the need to discard the progress made by the application. We explore two alternative recovery strategies, which use ULFM along with application-driven in-memory checkpointing. 
In the first case, the application is recovered with only the surviving processes, and in the second case, spares are used to replace the failed processes, such that the original configuration of the application is restored.
Our experimental results demonstrate that graceful degradation is a viable alternative for recovery in environments where spares may not be available.

\end{abstract}


\section{Introduction}
\label{sec:Introduction}

Ensuring resilient operation is a major design hurdle for current petascale high performance computing (HPC) systems as well as for future exascale systems. Analyses of operational system logs of current systems indicate a shrinking mean-time-to-failure (MTTF), a trend that is expected to get worse in future systems~\cite{Tiwari:SC17} with the emergence of highly complex HPC systems that contain heterogeneous, multi-component hardware and software.
For most long-running HPC simulations, the low MTTF means that the application may experience multiple failures during their execution.  
Although it is possible to design highly resilient systems based on hardware mechanisms, their costs and design efforts are prohibitive. Therefore, for many generations, HPC systems have used commodity hardware and software components to meet stringent timelines and budgets. However, the successful use of these systems in the presence of high fault rates requires software-based mechanisms to ensure their reliable operation. 

HPC applications are susceptible to both hardware and software faults and errors, which may be transient or permanent in nature. Transient or soft errors are random events caused by radiation particles affecting processing, network, or memory elements in the system~\cite{Shalf2011}. Permanent or hard errors are caused by wear-out or device aging effects, which manifest under specific conditions causing system components to malfunction. From the perspective of an HPC application, such transient and permanent errors may cause corruptions in the application data or computations, and on occasion cause fatal process crashes. 

For a parallel HPC application using message-passing based communication, the failure of a single process prevents forward progress of the overall application. The process failure causes the execution to cease and an abort signal to be sent to all surviving processes. HPC applications employ checkpoint and restart (C/R) mechanisms to recover from such failures. C/R solutions take periodic snapshots of the global system state, and upon a failure, the application is resumed using the latest checkpointed state rather than starting over. However, due to the scale of modern extreme-scale systems, global C/R is an increasingly inefficient strategy in the presence of very high failure rates.  

The recent proposal of user-level failure mitigation (ULFM) extensions~\cite{ULFM:IJHPCA} to Message Passing Interface (MPI) standard proposes primitives to repair the communicator and to enable the surviving processes to continue execution despite the failure of one or more MPI processes.
However, ULFM does not provide concrete failure recovery strategies, nor does it provide the ability to recover lost application state, leaving the user to decide which strategy to adopt for their applications. 
In this paper, we explore the design of two distinct in-situ recovery strategies in HPC applications using the MPI-ULFM interface. 
In the first approach, called \emph{shrink}, we isolate a failed process and continue execution with surviving processes, and in the second case, called \emph{substitute}, we restore the original design-time configuration of the application by use of spare processes. In contrast to prior work~\cite{Teranishi:MPI-UG-14} that focuses on performance evaluation of the ULFM primitives, we present a detailed evaluation of two complete failure recovery strategies. We evaluate our strategies for an application based on the widely used Generalized Minimal Residual (GMRES) linear solver.

The ULFM extensions to MPI do not provide the ability to reconstruct lost process state. While traditional checkpoint/restart solutions take a global system-wide snapshot, they tend to incur high storage, bandwidth and application performance overheads. In this work, we also develop an in-memory checkpointing solution to aid our ULFM-driven shrink and substitute strategies. This approach is application-driven, which facilitates a subset of total application state required for forward progress of an application to be preserved. Using point-to-point communication primitives, which are highly optimized in HPC systems, the checkpoints are stored in the memory of neighboring nodes in the system. This enables rapid recovery from process failures. Additionally, each node can assign multiple `buddy' nodes that can store its checkpoints, which supports handling multiple process failures.

The major contributions of this work are: 
\begin{itemize}
\item We present a detailed evaluation of process failure recovery strategies of \textit{substitute} and \textit{shrink} in the context of parallel applications.
Each of the solutions provide the ability to mitigate process failures via reconfiguration of the MPI communicator as well as application state recovery capabilities. 

\item We use an iterative linear solver application as a use case for demonstrating 
the various trade-offs and design complexities involved in the implementation of
the shrink and substitute recovery approaches. 

\item We evaluate the performance overheads for in-memory checkpoint and recovery operations at large scale in the presence of multiple independent process failures, which are becoming increasingly common in modern HPC systems. 
\end{itemize}

\section{Related Work}

Prior work in supporting resilience for MPI-based applications has explored strategies by extending the MPI runtime without changes to the interface. For example, MR-MPI \cite{Engelmann:2011}, rMPI \cite{Ferreira:2011:rMPI} and their successor RedMPI \cite{Fiala:2012} have introduced redundancy at different granularities. The key benefit of these runtime approaches is the avoidance of modifications to the application source code to support resilience.  

In contrast, ULFM requires programmer intervention, but provides much greater flexibility in developing failure recovery strategies and opportunities for optimization of the recovery process. Previous works have explored the viability of various strategies using ULFM to mitigate process failures for different applications.
For example, the Local Failure Local Recovery (LFLR)~\cite{Teranishi:MPI-UG-14} approach uses an early implementation of ULFM MPI to facilitate recovery using spare processes. For a finite element mini-app, which used checksums for application state recovery, the results showed that recovery time is dominated by the time to fix the communicator. Using a molecular dynamics application as case study, another study evaluated recovery by isolation of a failed MPI process by shrinking the communicator and excluding the failed rank~\cite{Laguna:MPI2014}. 
In a comparison of the post-recovery performance of ULFM-driven approaches using a synthetic benchmark~\cite{postRecovery:EuroPar15}, the performance of collectives after shrink operations was found to deteriorate, since MPI implementations commonly optimize process counts in terms of powers of two. Similarly, the work attributes the degradation of performance after process substitution to the distance of spare processes from failed node, which is strongly dependent on the topology of the HPC network.
However, a detailed comparison of these strategies for the same application on a uniform HPC platform to understand the design trade-offs and performance overheads has not been previously performed.

To improve the usability of ULFM, Fenix~\cite{Gamell:2014} provides a wrapper interface for applications to perform application-level in-memory checkpointing of variable state in addition to transparent state recovery assuming spare processes are present. 
A compiler-based solution was developed~\cite{CPPC:Journal-Supercomputing2017} to automatically identify safe code locations where a checkpoint can be consistently performed across all parallel processes. This solution optimizes the performance of ULFM-based recovery.

In this work, we utilize in-memory checkpoints rather than creating global checkpoints written using the parallel file system. Previous work has also explored various algorithm-based checkpoint-free schemes for application state recovery including the use of row and column checksums on dense matrix structures \cite{Huang:1984}, the use of additional dot products to detect errors in the matrix and vector elements of sparse matrix  computations \cite{Shantharam:2012}, etc. While these techniques may offer lower performance overheads, checkpoint-based recovery is more broadly applicable to various application codes.

\section{Background: Checkpoint/Restart}
\label{sec:Background}

Checkpoint restart (C/R) is a broadly applicable technique to mitigate process failures in distributed applications. 
It involves taking checkpoints and storing them redundantly so that if a failure were to occur, the application can recreate application state as it was prior to the failure. 
Different C/R strategies are possible depending on \textit{what} aspects of the application state are checkpointed, \textit{when} the checkpoints are created and \textit{where} the checkpoints are stored.
It is possible to store only specific data structures, or even the entire system state at the kernel-level or user-level. These two extremes represent a trade-off in terms of memory required to store the state and ease of implementation in terms of not needing to modify the application source code. Maintaining checkpoints of the entire system state across all processes can result in high storage overheads. On the other hand, preserving a limited amount of user-identified application state that is required for forward progress can significantly lower the memory overheads. 

When a checkpointing involves numerous processes, coordinated or uncoordinated strategies are possible. Coordinated checkpoints require that all parallel processes synchronize in order to ensure that there are no in-flight messages, which enables checkpointing of consistent global state. Uncoordinated checkpoints are created independently by each process. This requires message logging such that a consistent state can be recreated across all processes upon occurrence of a failure. However, a message sent from one process to another may be missed during the checkpointing operation and may need to be logged separately. Therefore, coordinated checkpoints are often preferred in practice. 

Among the key considerations of a C/R solution is the checkpointing frequency. If checkpointing operation is performed less often, the recovery from failures requires more time due to the need for recomputation from the point of the last stable checkpoint. When the checkpointing is performed too often, there is increased overhead to the application performance during failure-free operation. Therefore, the  trade-off between checkpoint interval and recovery latency must be carefully considered. The optimal checkpoint interval is dependent on the $MTTF$ of the system and checkpoint cost $C$. Young~\cite{Young:1974} identified the optimal checkpointing interval assuming an exponential failure distribution as: 
\begin{math}
  \sqrt{2*C*MTTF}
\end{math}

The checkpoint cost $C$ is strongly dependent on where the checkpoints are stored. In recent HPC systems, the parallel file storage system was used to store the checkpoints. This storage was assumed to be reliable, and could be used to retrieve the checkpoints in case of a failure. However, since this resource is shared by potentially multiple applications, and can generate localized congestion in the network, other alternatives have been explored. One alternative is to store the checkpoints redundantly in the local memory and in the memory of a buddy process. This configuration can be used to restore lost state in case one of the processes carrying the redundant state is alive. Optimized point-to-point message communication can be utilized for transferring checkpoints between buddy processes. 
This approach is possible with the use of application-assisted checkpointing, which can lower memory overheads. 
The performance efficiency and scalability of this approach has been demonstrated in prior works~\cite{SCR:2016:SSDs}. 
The overhead to transfer checkpoints to buddy processes can be eliminated by use of non-volatile memory, which is expected to be common-place in future systems. Multilevel checkpointing is also possible, where redundant checkpoints are maintained across multiple memory layers and checkpoint intervals for each layer are adjusted based on the cost of performing checkpoint at that layer~\cite{FTI:SC11}.

Therefore, the key factors in the design and implementation of a C/R solution are the scope of the state captured during the checkpoint operation, the frequency and the storage location for the checkpoint.
The primary overheads involved are the time to perform checkpoints and extent of recomputation during recovery, which are referred to as the \textit{waste} overhead of a C/R implementation. The main goal is to reduce this overhead. Other overheads associated with process failure tolerance include: overhead of detecting process failures, time to reconfigure and recover parallel runtime environment after a failure and the time to recover application state. The discussion of these factors is provided in the next sections.

\section{Detection, Reconfiguration and In-situ Recovery for Mitigation of Process Failures}

HPC applications rely on a distributed runtime system software for detection of process failures. 
We utilize the fault tolerant version of MPI communication library ULFM for reliable notification of process failures.
The ULFM implementation extends the MPI operations with capabilities to notify the application of anomalies in the group of processes involved in a communication operation, i.e., a process failure is notified when a MPI operation can not be completed as intended.
The fault detection capabilities of ULFM need to be initialized by changing the default error handler \code{MPI\_ERRORS\_ARE\_FATAL} on the communication objects. During the execution of a MPI routine, if an error occurs, a notification is raised to the processes involved in a communication operation via the return code \code{MPI\_ERR\_PROC\_FAILED}. Consensus-based algorithm among neighbors and timeouts are commonly used to detect failed processes.
These detection mechanisms incur some overhead in comparison to the standard MPI implementations that do not provide these features.
However, recent efforts show considerable decrease in these overheads and scalable operation~\cite{Bosilca:2016-SC16-ULFMdetect}.

From an application's perspective, the error reporting capability of ULFM provides the ability to take either a \textit{proactive} or a \textit{reactive} approach to failure detection.
If an application requires timely notification of failure to all processes, then a proactive approach may involve strategic placement of collective operations inside the code, such that a failure is detected early on and costly re-computation is avoided. However, frequent calls to collective routines introduce high synchronization overheads.
Therefore, another approach is to wait for error notification by a MPI operation in the code, i.e., a reactive approach. 
This can involve checking the error codes of every MPI call, or at least a selected subset of calls. This may also be easily accomplished by implementing a MPI error handler that is called every time an error is notified by one of the MPI routines. Error propagation to other processes and recovery is orchestrated by this error handler. 

Once each surviving process has been notified of the failure in the parallel computing environment, then reconfiguration and recovery needs to be performed in a manner which is most beneficial to the HPC application.
To begin with, we utilize ULFMs capability to remove all failed processes from communication objects (by using \code{MPI\_COMM\_SHRINK()}). The generated pristine MPI objects are used in future communications among parallel processes.
Afterwards, it is left on to the user to recover application state and to resume forward progress of the application towards the final solution. 
In doing so, we have two options, either resume with same number of processes as the application began with, a strategy referred to as \textit{substitute}, or resume with reduced number of processes, a strategy referred to as \textit{shrink}. 
We explore the implications of both these options on application assisted in-memory checkpointing in this section. Both have different design tradeoffs, and their impact on application performance will be evaluated in the experiment and results section. 

\begin{figure} [tp]
\centering
\includegraphics[width=0.8\linewidth]{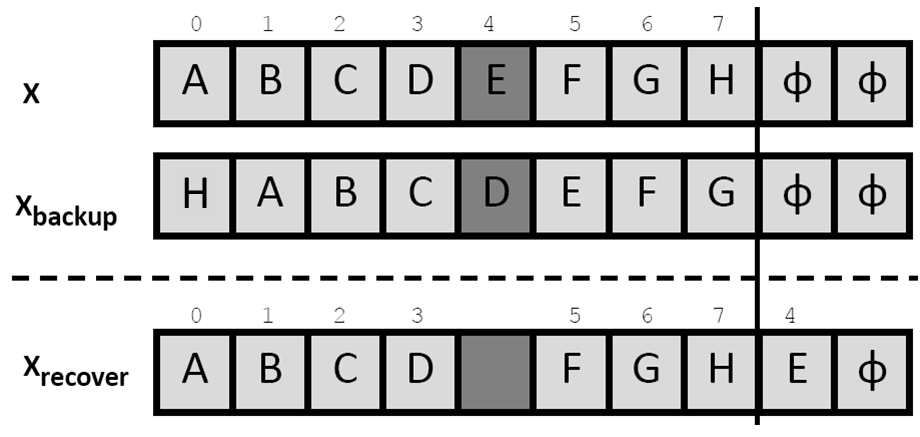}
\caption{Substitute approach: recovery mechanism for both static and dynamic distributed objects. The spares are represented by $\phi$.}
\label{Fig:WarmStandby}
\end{figure}

\subsection{Substitute: Supplemental Computation with Spares}
The \textit{substitute} approach requires allocation of spare processes which can take the form of warm or cold spares, depending on whether the processes are allocated at design time or spawned at runtime. 
In this work, we do not consider hot spares since those entail redundant computation for each and every process, e.g., modular redundancy-MPI~\cite{Engelmann:2011}. MR-MPI employs complete state replication across spares and therefore does not require checkpointing of application state. 
This can be extremely resource intensive yet sudden degradation in reliability is incurred once failures start to happen since failed processes are not replaced and subsequent failures in non redundant processes can be fatal.
Therefore, we resort to spares which can be integrated into the application as the need arises, i.e., as failures occur. 
The spares which are design time allocated are warm spares, and processes spawned at runtime are referred to as cold spares. 
Use of both approaches is appropriate and similar from the context of the application, but spawning processes at runtime has more overhead. In addition, some computing environments restrict spawning of new processes by an already scheduled job. 
We therefore discuss the use of warm spares for restoring the original configuration of the application, as a substitute strategy. One obvious disadvantage in this case is the non-utilization of resources in the failure free case. 

The availability of spare processes provides the opportunity for processes to continue execution with workload similar to that of a failure-free setup. 
This is extremely useful for applications which perform initial compute-intensive data distribution based on the input.
For example, balancing the number of non-zero elements assigned to each process in a sparse input matrix requires the use of graph processing algorithms. 
Additionally, the use of spare strategy is mandatory in some applications due to constraints on problem decomposition, e.g., cube number of processes are required if the problem is being decomposed onto a cubic mesh. 

On the implementation side, programming effort is required to integrate the spare processes into the application code. 
The effort involved is highly dependent on the programming language used and the structure of the code if an existing code base needs to be made resilient.
The allocated warm spares need to be segregated at the beginning of the computation, and wait for their utilization during failure recovery.
Once pristine communicator objects have been attained, the spare process can be stitched in. 
In the new configuration, the spare process is assigned a rank/id (the ranks for processes are represented by unique numbers) similar to that of the failed process as shown in Figure~\ref{Fig:WarmStandby}.

After we fix the parallel runtime environment and its objects, the application state of the spare process needs to be populated. 
We use the checkpoints taken prior to a failure to construct the application state for the spare process.
The spare process communicates with the neighbor of the failed process for this purpose since it maintains a copy of application state for the failed process as shown in Figure~\ref{Fig:WarmStandby}. In Figure~\ref{Fig:WarmStandby}, the portion of application state for each process is represented by letters A, B, etc., and the backup data-structure illustrates how a redundant copy is maintained at the neighboring process. 
During recovery, the survivors use their local copy to restore the application state while the spare process uses MPI point-to-point communication to get its portion of the application state from the backup data-structure. 
Along with the recovery of distributed state space, there is also a need to synchronize the state of the processes which is local to them. 
This is supposed to be consistent across processes, and we can use any surviving process to populate the local state of the spare process. 
For example, the number of iterations across all processes should be same, otherwise, the spare process may diverge and possibly cause a deadlock.
Thus, this is a crucial step in the application state recovery process.

\begin{figure} [tp]
\centering
\includegraphics[width=0.8\linewidth]{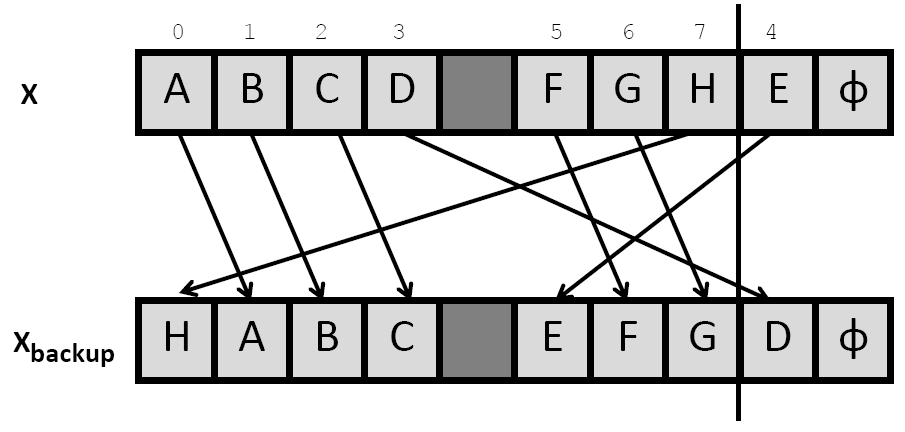}
\caption{Checkpointing operation in case of substitute approach after utilization of a spare process.}
\label{Fig:standbyRecovery}
\end{figure}

The application can resume after consistent recovery of application state across all processes. 
The point of restart needs to be coordinated among all processes. For example, exception handling blocks can be used to jump to the start of an iterative block after recovering from a process failure. 
To sustain future failures, the application continues to make checkpoints of dynamic state across the set of processes involved in the computation as shown in Figure~\ref{Fig:standbyRecovery}. 
There is no change in the size of checkpoints taken across these processes as compared to failure free case. 
However, since spares may be placed on a physically distant node, assuming the mapping is not changed during the computation. An increase in latency can be observed during ordinary communication operations as well as during the checkpoint operation. 
For instance, a neighboring process based on rank may not be a physical neighbor after the spare processes have been utilized. This can give rise to arbitrary communication patterns instead of regular communication patterns which may be present in the application by design. 
This effects performance during message communications depending on the layout of the HPC network. 

\subsection{Shrink: Graceful Degradation with Survivors}
The \textit{shrink} strategy for process failure recovery alleviates the need to have spare processes allocated at design time. 
However, determining the right number of spare processes can be challenging and may depend on multiple factors, such as the number of failures expected (based on MTTF of the system) and the number of available resources. 
On the other hand, arbitrary number of process failures can be sustained using the shrink approach, as long as there are enough surviving processes to continue execution of the application without significant performance impact.
In this approach, all processes get to perform useful work from the start. 
However, good performance or throughput is strongly dependent on whether the application can dynamically adjust the workload across all the surviving processes. 
Non uniform workload distribution can affect performance significantly since some processes may end up doing bulk of the work and overall application performance is determined by the slowest process. 

\begin{figure} [tp]
\centering
\includegraphics[width=0.8\linewidth]{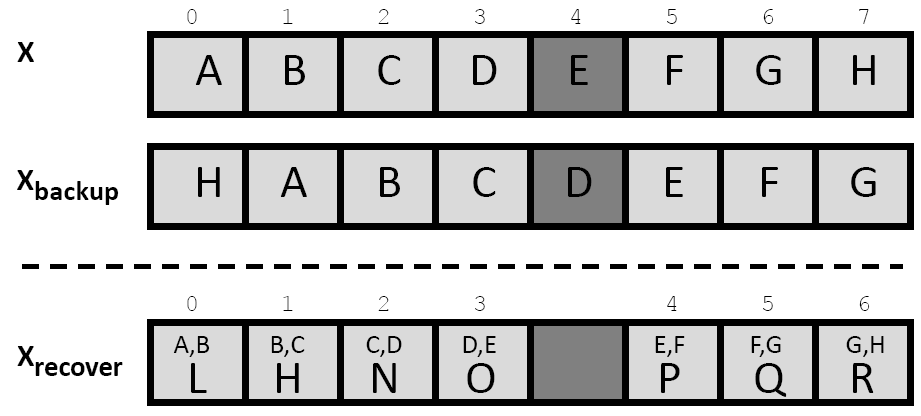}
\caption{Shrink approach: checkpoint and recovery mechanism for static and dynamic distributed objects.}
\label{Fig:Shrink}
\end{figure}

The recovery of application from process failures in this case involves  effort from the user.
This is because domain knowledge is required in most cases to re-distribute workload among surviving processes. 
To illustrate, let us take the simple example of distributing a vector with $R$ rows among $P$ parallel processes. Assuming perfect divisibility, each process gets $\frac{R}{P}$ rows in a block wise manner in the beginning. 
After the failure of one process, we need to redistribute the $R$ rows among $P-1$ processes. 
For uniform workload distribution, each surviving process needs to be assigned at least $\frac{R}{(P-1)} - \frac{R}{P}$ extra rows.
This is accomplished via inter-process communication among neighboring processes. 
The cycle is repeated every time a failure occurs and the workload on each survivor keeps on increasing as more failures are encountered. 

The above re-distribution can take place during state recovery while using the checkpointed dynamic and static state of the application as shown in Figure~\ref{Fig:Shrink}. In-memory checkpoints with redundant data across local and remote processes ensures that every process get its desired chunk of the assigned data. In some cases, it is possible to exclusively use local data to reconstruct desired state. 
For the example illustrated in Figure~\ref{Fig:Shrink} whereby process with rank 4 (process id) has failed. All processes with rank less than 4 need to communicate with their neighbors to get a chunk of their local data.
For instance, the process with rank 3 needs to communicate with process having rank 5 to grab a chunk of its backup data.  
Whereas, the processes with ranks greater than the rank of the failed process can use their local data alone to conform to the new data distribution pattern.
Consequently, the amount of communication during state recovery depends on the location of the failed process, i.e., failure of processes with higher ranks results in more messages on the network.

After the re-distribution and recovery of application state, we need to update all the in-memory checkpoints. 
This helps to ensure that future failures are sustained and every distributed data chunk has a backup. 
This adds on to the cost of state recovery. 
Any future checkpoints of dynamic state must comply with the new distribution plan.

We do not need to recover local variables in case of shrink strategy and the application can resume execution after the above steps. 
With this approach, the number of workers decrease over time and checkpoint overhead per process increases over time. 
It is imperative to consider all these overheads while choosing a recovery strategy. 

\section{Use Case: Iterative Solver}

Iterative methods for solving linear equations assume reliability of data and arithmetic operations. If faults occur during execution, they cause the solver to either abort or to compute an incorrect result with no warning.  
To support fault resilience, application-level solutions often rely on algorithm-level techniques that encode computations using linear error correcting codes, or leverage the mathematical convergence properties of the solver to ensure it produces a correct outcome at the cost of needing additional iterations. 

A fault tolerant version of the Generalized Minimal Residual (GMRES) called FT-GMRES \cite{Hoemmen:2011} uses the notion of selective reliability to ensure that the solver converges to a correct outcome. The original GMRES method was developed as a Krylov subspace method for an iterative solution of large sparse nonsymmetric linear systems of the form $Ax = b$ \cite{Saad:1986}. The method was improved by partitioning the solver into inner-outer iterations, where the ``inner" solve step preconditions the ``outer" flexible iteration \cite{Saad:1993}. The FT-GMRES algorithm requires that only the outer iterations are highly reliable, while any faults during the execution of the inner iterations are tolerated. This partitioning of the solver, which executes only a fraction of the solver's computations in highly-reliable mode, offers protection against possibly unbounded numerical errors caused by silent data corruptions.

However, MPI-based applications that use the FT-GMRES method are not protected against hard errors that may cause one or more of the MPI ranks to fail. Such process failures caused by hardware component malfunctions or compute node failures are unavoidable in large-scale HPC platforms.

\section{Implementation and Experimental Setup}
\label{sec:ExperimentalEval}

\textit{Implementation details:}
In this work, we have utilized the FT-GMRES solver from~\cite{Elliot:FTGMRES}, which is implemented using C++ based Tpetra package within the Trilinos 12.6.4 framework~\cite{Trilinos}.
The Tpetra package provides the ability to distribute large objects such as sparse matrices, dense vectors onto multiple parallel processes using a variety of parallel programming models. 
It allows FT-GMRES to do parallel operations such as sparse matrix vector multiplication, vector scalar multiplication, etc.  
Using this setup, FT-GMRES is able to solve large-scale linear systems. 

Our contributions to FT-GMRES include:
1) altering the application to support MPI-ULFM instead of standard MPI,
2) adding process failure detection and reconfiguration mechanisms, 
3) adding support for in-memory checkpoints to recover application state,
4) adding dynamic workload redistribution in case of shrink strategy, and
5) adding warm spares to be utilized for recovery in case of substitute strategy. 

In this work, we have utilized the ULFM version 1.1 which uses Open MPI version 1.7.1 as a base implementation.
All necessary changes are made to the code for integration of MPI-ULFM and to leverage its capability to resume application despite process failures. 
Specifically, ULFM is able to detect process failures reliably, identify the identity of the failed processes, notify survivors about process failures, and reconstruct the communication objects.
Process failures are detected proactively by a custom MPI error handler, i.e., control is transferred to the error handler whenever an error code is returned by any MPI call. After recovery is complete, we leverage C++ exception handling to coordinate a uniform restart location across all processes, i.e., we jump to beginning of the iterative block. 

Our implementation provides the ability to checkpoint Tpetra objects of sparse matrix and dense vectors. 
These objects are checkpointed in the memory of buddy nodes according to user-defined mapping and checkpoint intervals. 
For instance, dynamic objects such as the solution vector which changes at every iteration of the solver is checkpointed according to user defined interval, whereas, static objects such as matrix $A$ and right hand side vector $b$ only need to be checkpointed upon a process failure. 

\textit{Evaluation platform:}
We use a 960-core Linux cluster with a fully connected dual-bonded 1 Gbps Ethernet for our experiments. 
Each compute node has 2 AMD Opteron processors each (with 12 cores each) and 64 GB memory. 
The interconnect supports non-blocking point-to-point bandwidth of 215 MB/s.
We use process counts of 32, 64, 128, 256, and 512 in our experiments. The mapping of processes onto cores is designed to incorporate the effect of communication over the network. 

\textit{Test problem:}
A test problem is generated by discretizing a regular 3D mesh in Trilinos framework. 
We solve a linear system with sparse matrix $A$ having about 7 million rows and 186 million nonzero elements.
We fix the problem size for all experiments. 
Thus, the number of elements assigned to each process decreases as we increase the scale of our experiments.
In failure free setup with above problem configuration, the solver converges to a solution within 325 iterations total. 

\textit{Process failure injection:}
We perform controlled process failure injections in our experiments to produce reproducible results as follows:
(1) The rank positions of failed processes are fixed throughout the experiments. 
These positions are selected to represent worst case scenarios for shrink and substitute approaches.
In case of substitute recovery experiments, the failed process is selected to be on a different physical node from the node on which the spare processes reside. 
By default, the spare processes are mapped to the later nodes in the experiment. This ensures network communication delays are added when spares are utilized by the application. 
Whereas, process failures are injected towards higher ranks when shrinking recovery strategy is used, as it represents the worst case in network communication during state recovery phase as described earlier (see Figure~\ref{Fig:Shrink});
(2) The failure injection time window is fixed for each injection experiment.
Note, we checkpoint dynamic state only after the completion of one inner solve operation (every 25 iterations of the solver), thus there is an upper bound on the amount of re-computation.

During our experiments, we inject up to four independent process failures to simulate the effects on long running HPC simulations.
We therefore evaluate the sustainability of both recovery approaches.
It helps us to understand how the recovery characteristics or overheads change from one failure to another. 
It also provides us the opportunity to model these overheads in terms of a single process failure overhead, which is useful since it alleviates the need to run extra experiments.
In each case, a process failure injection is simulated by a system call to \code{SIGKILL}.
Furthermore, we assume the presence of adequate number of spares when utilizing the substitute recovery strategy.

\section{Experimental Results}
\label{sec:Results}

\begin{figure} [tp]
\centering
\includegraphics[width=\linewidth]{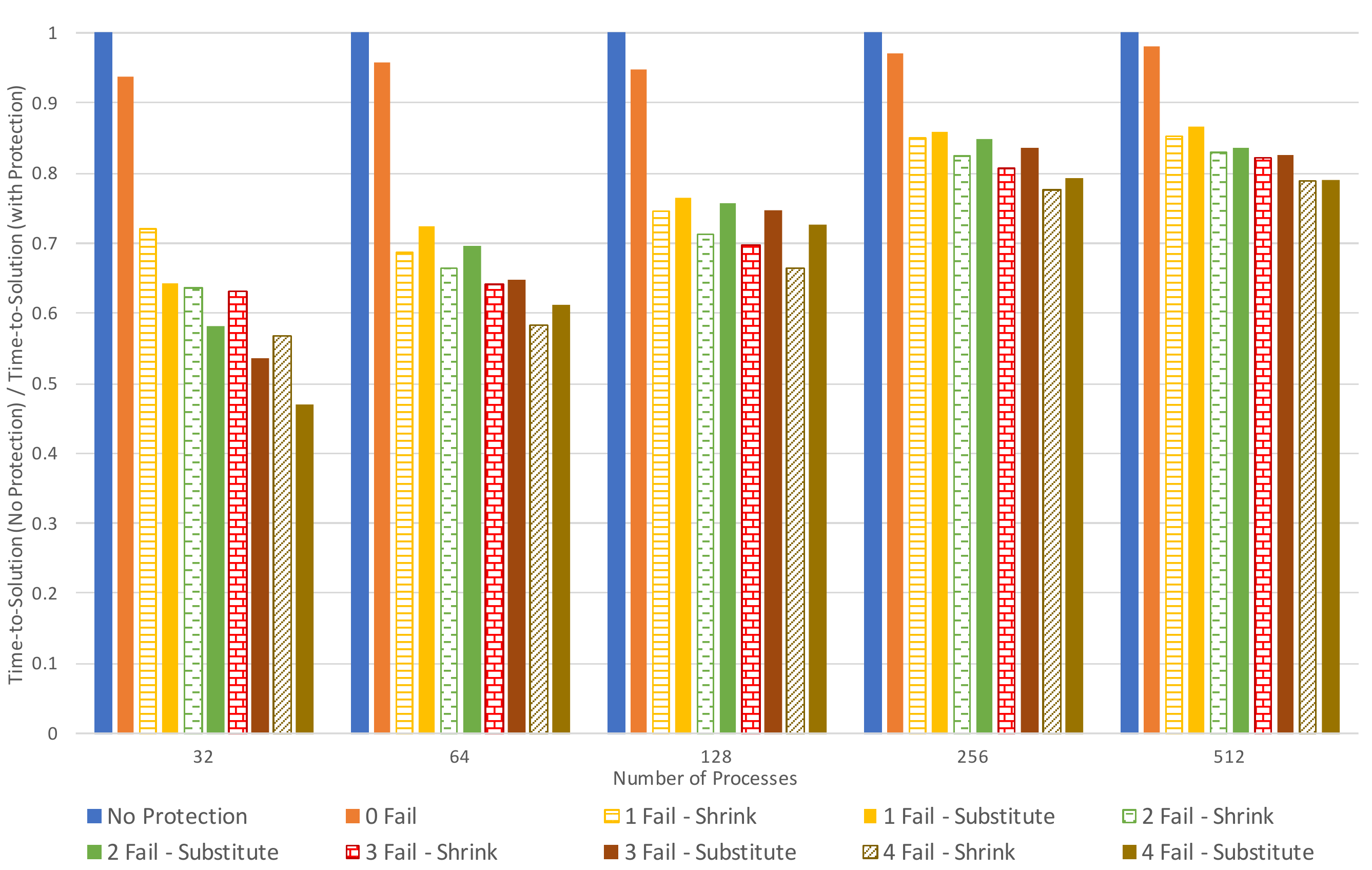}
\caption{Performance comparison between the shrink and substitute recovery strategies with up to four process failure injections.}
\label{Fig:HEonlyOverall}
\end{figure}

In this section, a detailed performance comparison of shrink and substitute approaches for in-situ recovery to process failures is done for the FT-GMRES solver.
Figure~\ref{Fig:HEonlyOverall} demonstrates the performance slowdowns or reduction in time-to-solution with each recovery approach sustaining multiple independent process failures in comparison to the no protection case.
Each point in this plot is obtained by averaging results from multiple experiments such that standard deviations are low, i.e., the coefficient of variation range between 0.01 and 0.15. 
Most of the variation in results is due to the recomputation overheads since the position and window of failure injection is fixed in all experiments, as discussed earlier.
The bars with patterns in Figure~\ref{Fig:HEonlyOverall} represent the performance of shrink strategy and solid bars represent the performance of substitute strategy. 
The values close to one (no protection) translate to a lower overhead. 
For instance, the `0 Fail' case shows the cost of providing process failure tolerance in case of failure free conditions.
Results demonstrate that the overheads to perform the checkpoint decreases with scale since the number of elements to checkpoint per process decreases.

\begin{figure} [tp]
\centering
\includegraphics[width=\linewidth]{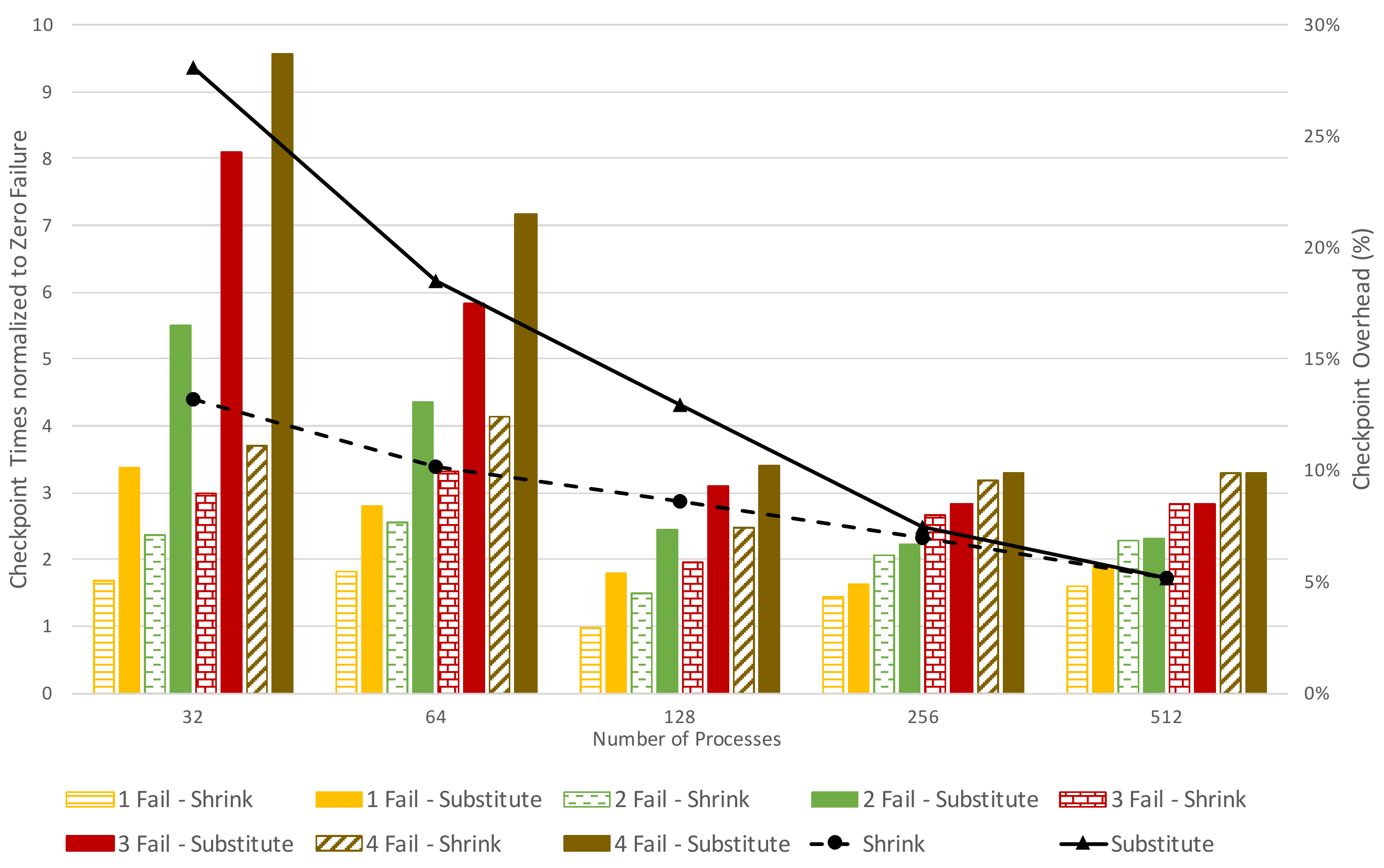}
\caption{Checkpoint times normalized to the no failure case.}
\label{Fig:BackupMultiple}
\end{figure}

A significant slowdown is observed in case of both shrink and substitute approaches when mitigating process failures. The overhead increases with increasing number of process failures since the state recovery overheads are additive. 
Recall that the performance overheads when tolerating process failures include the checkpointing overheads, reconfiguration overheads, state recovery overheads, and re-computation overheads. 
The use of spare processes provides a performance advantage as compared to the shrink strategy at process counts greater than 32 for all injection campaigns.
However, this advantage starts to diminish with increase in scale since there is not a substantial increase in workload at each process when using the shrink approach.
A more thorough analysis of the overheads in each case reveals that the checkpoint overheads tend to be higher for the substitute recovery strategy as shown in Figure~\ref{Fig:BackupMultiple}, especially at lower process counts. 
Results demonstrate that the checkpoint overheads do not increase multiplicatively with increasing failures for the substitute recovery approach. 
We attribute this behaviour to the placement of spares as discussed later on.
On the contrary, a linear pattern is observed in case of shrink recovery approach. 

Figure~\ref{Fig:BackupMultiple} also shows the comparison of checkpoint overheads with respect to the total time to solution between shrink and substitute approaches with four process failures plotted on the secondary y-axis.
It can be observed that the checkpoint overhead for substitute approach is high as compared to the shrink approach at processor counts of 32, 64 and 128, and is comparable at processor counts of 256 and 512.
The overhead is as high as 28\% at a processor count of 32 and tends to decrease to as low as 5\% at a processor count of 512.
At large scales, the checkpoint overheads are similar for both approaches due to smaller inter-process communication overheads. 
On the other hand, the observed difference in overheads at small scale is due to placement of spares in the system.
Since we use fixed mapping throughout the computation as provided in the standard implementation of MPI, the communication with utilized spare processes tend to add to the overheads. 
The spare processes are always mapped towards the later nodes (highest ranks are assigned to the spares).
To aggravate this issue, the process failures in our experiments are forced to be on a different node as compared to the location of the spares. 
Therefore, when the spare processes replace the failed processes, it can lead to higher communication overheads as discussed earlier (see Figure~\ref{Fig:standbyRecovery}). 
This overhead is increased when we use more spare processes, i.e., more failures are mitigated. 
We expect this overhead to be present in other parallel operations as well.

\begin{figure} [tp]
\centering
\includegraphics[width=\linewidth]{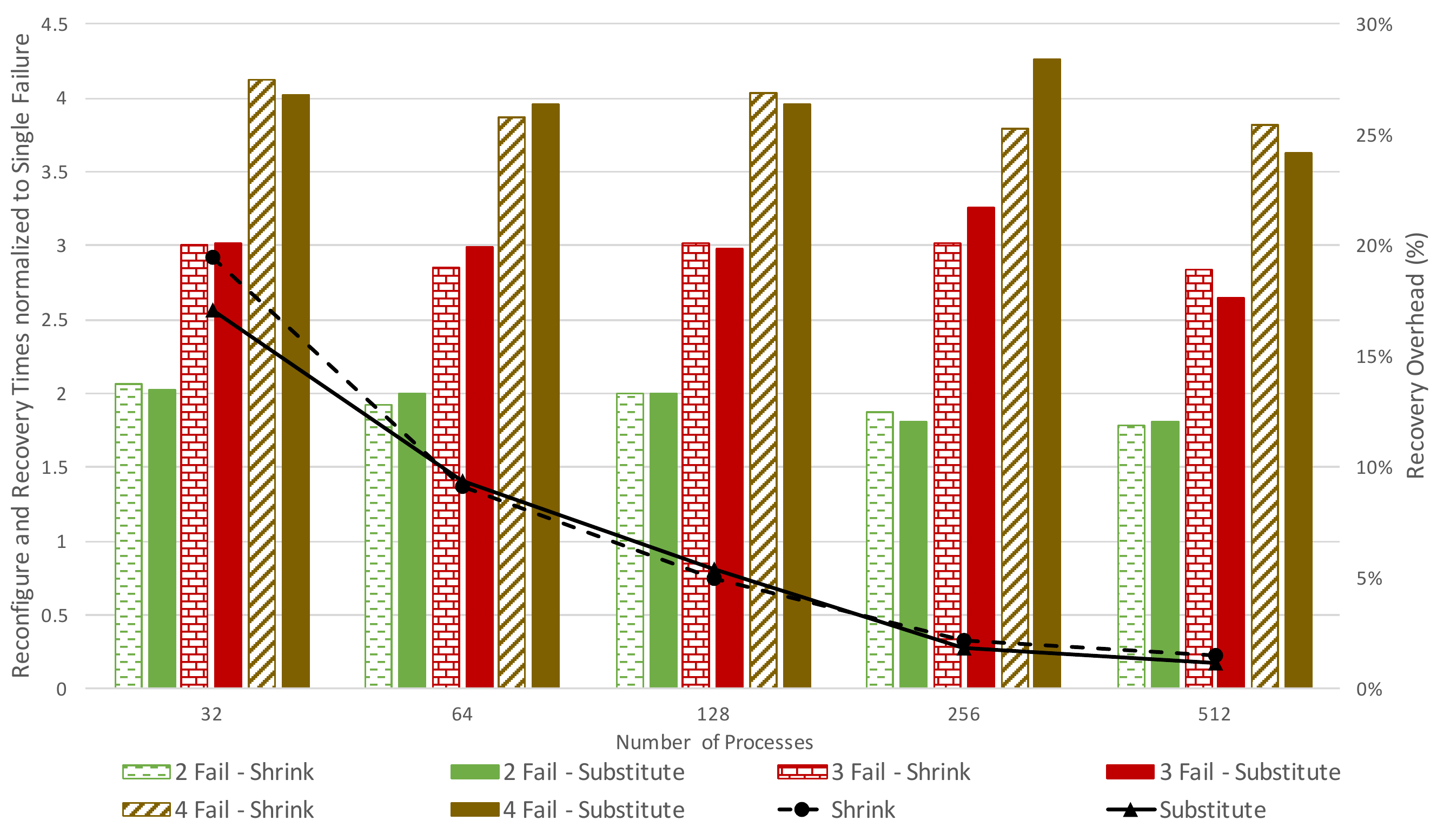}
\caption{Recovery and reconfiguration times normalized to single process failure.}
\label{Fig:RecoverMultiple}
\end{figure}

A comparison of recovery overheads between shrink and substitute approaches is shown in Figure~\ref{Fig:RecoverMultiple} (the overheads w.r.t. total time to solution are plotted on the secondary y-axis). 
For most cases, the recovery overheads tend to be comparable and range from 19.5\% to as low as 1.5\%. 
Note, there is a drop from 19.5\% to 9\% going from 32 to 64 processes. 
Figure~\ref{Fig:RecoverMultiple} also shows that it is relatively straightforward to estimate the overheads for multiple failures from the recovery costs of a single failure.
Almost similar recovery overheads in both cases show that most of the overhead is due to inter-process communication required to reconstruct application state. Therefore, the workload re-distribution overheads in case of shrink approach tend to be negligible. 
Our results also indicate that the reconfiguration overheads are negligible, i.e., they range from 0.01\% to 0.05\%. Although, a slight increase in overhead is observed in our experiments when the spare process needs to be stitched into the repaired communicator object. 
Overall, both recovery and reconfiguration overheads tend to be lower as compared to the checkpoint overheads.

\textit{Discussion:} Our experimental results show that the shrink approach provides graceful performance degradation at large scale when enough workers are present to share the workload of the failed processes. 
This is based on the assumption that workload redistribution is supported in the application which is observed to have negligible overhead in our experiments. 
Our results also show that the mapping of spare processes can significantly affect application performance especially at smaller scales since it disrupts regular communication pattern of the application. 
This can mitigate the performance benefit of having spares when communication overheads in the application are dominant. 
Finally, the results presented are widely applicable for large-scale high-end scientific applications which support parallelism by distributing matrices and vectors among processes.
The Tpetra package, which has been modified for this work is widely used in the HPC scientific community. 

\section{Conclusion}
\label{sec:Conclusion}

In extreme-scale HPC systems, various types of malfunctions and component failures occur at very high frequencies. For parallel HPC applications developed using MPI, these events are often fatal since the failure of even a single process causes the remaining processes in the MPI communicator to block indefinitely, preventing forward progress of the HPC application. While recent ULFM extensions to MPI provide simple primitives to support MPI communicator recovery, they don't explicitly support the recovery of lost application state, nor do they provide well-defined application recovery models.  
In this paper, we explored two alternative strategies for handling process failure recovery of MPI applications. We evaluated how the different implementations use application-driven process recovery with in-memory checkpointing to offer different levels of performance and scalability. We demonstrated how these strategies may be flexibly applied on an application-specific basis to gracefully handle failures
while minimizing any degradation in the application performance.

\section*{Acknowledgments}
The authors would like to thank James Elliot from Sandia National Laboratories for his help with the FT-GMRES code. \\
This material is based upon work supported by the U.S. Department of Energy, Office of Science, Office of Advanced Scientific Computing Research, program manager Lucy Nowell, under contract number DE-AC05-00OR22725.

\bibliographystyle{IEEEtran}
\bibliography{IEEEabrv,references} 

\end{document}